\documentclass[useAMS,usenatbib]{mn2e}
\usepackage{epsfig,lscape,graphicx,times}

\DeclareGraphicsExtensions{.eps,.eps.gz,.epsi}
\newcommand\Te{$T_{\rm e}$}%
\newcommand\hiir{H~{\sc ii} regions}%

\title[Oxygen and neon problems]{Are oxygen and neon enriched in PNe and is the current solar Ne/O abundance ratio underestimated?}
\author[W. Wang \& X.-W. Liu]{W. Wang$^1$\thanks{E-mail: wangwei9527@gmail.com}
and X.-W. Liu$^{2,3}$\\
$^1$National Astronomical Observatories, Chinese Academy of Science, Beijing 100012, P.R. China\\
$^2$Department of Astronomy, Peking University, Beijing 100871, P. R. China\\
$^3$Kavli Institute for Astronomy and Astrophysics, Peking University,
Beijing 100871, P. R. China}

\begin{document}
\date{Received:}
\pagerange{000--000} \pubyear{0000}

\maketitle

\label{firstpage}

\begin{abstract} {

A thorough critical literature survey has been carried out for reliable
measurements of oxygen and neon abundances of planetary nebulae (PNe) and
\hiir. By contrasting the results of PNe and of \hiir, we aim to address the
issues of the evolution of oxygen and neon in the interstellar medium (ISM) and
in the late evolutionary phases of low- and intermediate-mass stars (LIMS), as
well as the currently hotly disputed solar Ne/O abundance ratio. Through the
comparisons, we find that neon abundance and Ne/O ratio increase with
increasing oxygen abundance in both types of nebulae, with positive correlation
coefficients larger than 0.75. The correlations suggest different enrichment
mechanisms for oxygen and neon in the ISM, in the sense that the growth of neon
is delayed compared to oxygen. The differences of abundances between PNe and
\hiir\, are mainly attributed to the results of nucleosynthesis and dredge-up
processes that occurred in the progenitor stars of PNe. We find that both these
$\alpha$-elements are significantly enriched at low metallicity (initial oxygen
abundance $\la$ 8.0) but not at metallicity higher than the SMC. The fact that
Ne/O ratios measured in PNe are almost the same as those in \hiir, regardless
of the metallicity, suggests a very similar production mechanism of neon and
oxygen in intermediate mass stars (IMS) of low initial metallicities and in
more massive stars, a conjecture that requires verification by further
theoretical studies. This result also strongly suggests that both the solar
neon abundance and the Ne/O ratio should be revised upwards by $\sim$0.22~dex
from the Asplund, Grevesse \& Sauval values or by $\sim$0.14~dex from the
Grevesse \& Sauval values.}

\end{abstract}


\section{Introduction}

It is generally assumed (e.g. \citealt{henry1989}) that $\alpha$-elements, such
as oxygen and neon, are not altered significantly by the various
nucleosynthesis and dredge-up processes that occurred during the late
evolutionary stages of LIMS, which range in mass from about 0.8 to
8\,$M_{\sun}$. Therefore, the abundances of $\alpha$-elements measured in the
descendants of LIMS, such as Asymptotic Giant Branch (AGB) stars and PNe,
should reflect the metallicity, $Z$, of the ISM, from which the parent stars
were formed. Under the assumption, abundances determined for PNe have been
widely used to constrain the chemical evolution of the Milky Way and of nearby
galaxies.

Recently, there is however some observational evidence that the above
assumption is questionable, especially in low $Z$ galaxies, where it has been
found that oxygen abundances of PNe are higher than the average value of
\hiir\, (e.g. SMC and LMC; Leisy \& Dennefeld 1996, 2006). The enrichment of
oxygen in PNe in lower metallicity environments becomes more apparent by
comparing the Magellanic Clouds and the Galaxy. On the other hand, oxygen
destruction has been found in PNe evolved from massive progenitors (e.g. NGC
6302; \citealt{pottasch1999}; we define the progenitor stars which were
initially more massive than about 5\,$M_{\sun}$ as massive progenitors, where
the ON cycle begins to occur). There is also evidence that asymmetric PNe may be
enriched in neon, as suggested by the observed relationships between the
abundances and morphological types of PNe (\citealt{corradi1995};
\citealt{stanghellini2000}). 

The various nucleosynthesis [e.g. the Hot Bottom Burning (HBB)] and dredge-up
processes occurred in the late evolutionary epochs of LIMS, which eventually
determined the chemical composition of the ejected nebulae. These depend
strongly on the initial metallicity as well as mass. Recent theoretical
calculations have indeed revealed several channels of altering oxygen and/or
neon abundances by some processes that occurred in the PN progenitors. Oxygen
destruction, for example, can occur through the CNO cycle specifically via the
ON cycle in massive progenitors, while the 3rd dredge-up can efficiently modify
the nebular oxygen abundance by transporting the freshly manufactured material
to the stellar surface. \cite{charbonnel2005} emphasized that the surface
abundance of oxygen in massive progenitors is a result of competition between
the efficiencies of the 3rd dredge-up (for production) and the HBB (for
destruction). We note that neon production may also become significant for a
narrow range of stellar mass (\citealt{KL2003}). 

One of the purposes of the current study is to clarify whether oxygen and neon
abundances are significantly modified in LIMS and if so, the modifications that
affect our current understanding of the chemical evolution of galaxies. To
address this issue, we have collected all recent reliable determinations of
oxygen and neon abundances of PNe and \hiir\, and present a thorough analysis
of available data in this paper. 

As a consequence of the application of a time-dependent 3D hydrodynamical model
of the solar atmosphere in place of the earlier 1D model, the metal content in
the solar convection zone has been reduced by almost a factor of two. Amongst
them, abundances of key elements C, N and O have been lowered by approximately
0.2--0.3~dex from the earlier widely adopted values (see a review by
\citealt{asplund2005}; henceforth AGS05). However, those large downward
revisions have caused serious conflicts with other solar physics studies. To
fit the helioseismological measurements, for example, \citet{bahcall2005}
suggested that the neon abundance needs be raised from 7.84 to 8.29 or the Ne/O
ratio from 0.15 to 0.42. Recent Chandra measurements of 21 nearby solar-type
stars by \cite{drake05} concluded that the solar Ne/O ratio should be around
0.41. Finally, observations of Galactic PNe yield an average Ne/O ratio of
about 0.25 from collisionally excited lines (\citealt{WL07}). As we shall show
in Section~3, the depletion or enrichment of oxygen in PNe is insignificant at
solar metallicity. Thus accurate determinations of oxygen and neon abundances
of PNe and \hiir\, can be used to constrain and calibrate solar oxygen and neon
abundances and their ratio. This serves as the goal of the current paper. 

The paper is organized as follows. In Section 2, we present our critical
literature survey. In Section 3, we compare the oxygen and neon abundances in
PNe and in H~{\sc ii} regions and discuss their possible evolution mechanisms
and histories. In Section 4, we discuss the neon abundances and the Ne/O ratio
in the Sun. In the last Section, we conclude by summarizing the main results.

\section{The data}

\cite{henry1989} found a tight linear correlation between the Ne/H and O/H
abundances for the Galactic, Magellanic and M\,31 PNe. Similar results have
also been obtained by \cite{henry2004} for a sample of 86 Galactic PNe, by
\cite{stanghellini2006} for a sample of 79 Galactic PNe and by \cite{WL07} for
a sample of 25 Galactic bulge PNe and 58 disk PNe. A linear regression for the
four samples yields slopes of 1.16$\pm$0.04, 1.18$\pm$0.095, 1.14$\pm$0.09 and
1.21$\pm$0.08, respectively. The tight correlation has been interpreted as a
direct consequence of the fact that both neon and oxygen originate from primary
nucleosynthesis in massive stars (MS, $M\ga 10 M_{\sun}$) and are therefore
independent of the evolution of LIMS, the progenitors of PNe. On the other
hand, it somehow seems strange that all those slopes are consistently larger
than unity, implying that the growth of oxygen and of neon may not occur
concurrently or at the same rate. \cite{WL07} further examined this important
issue by calculating the average Ne/O abundances of the SMC and LMC PNe, using
the data published by \cite{LD06}. The average Ne/O ratios of 37 SMC PNe and
120 LMC PNe are 0.154 and 0.183, respectively, significantly lower than the
corresponding values of 0.24 and 0.25 for the Galactic disk and the bulge
samples, respectively (c.f. Section~{7.3} in \citealt{WL07}). The mean oxygen
abundances of PNe in the SMC, LMC, Galactic disk and bulge are 8.09, 8.38, 8.60
and 8.70, respectively. It is clear that the Ne/O ratio and the oxygen
abundance are positively correlated.

To further explore this critical issue, we have carried out a thorough
literature survey of available measurements of oxygen and neon abundances of
PNe and H~{\sc ii} regions. The results are listed in Table~\ref{tab01} and
plotted in Fig.~\ref{fig01}. In addition to the Local Group galaxies, a sample
of nearby dwarf irregular galaxies (dIrr) studied by \citet{vanzee2006a} and a
sample of blue compact galaxies (BCD) by \citet{guseva2007} are also included.
For galaxies with more than one available reference, the one with the largest
number of objects or the most reliable data is used. The name of the host
galaxy and the region sampled, the number of objects in each sample, the mean
Ne/O, O/H and Ne/H values and the data reference are given in Columns 1~--~6 of
Table~1, respectively. For each sample, only those objects with electron
temperature (hereafter, \Te) measured from the direct method are included, to
ensure the reliability of the abundance determinations. The numbers given in
parentheses are the corresponding standard errors of the mean, except for
samples consisting of a single object, for which the measurement errors are
given instead. For the Orion Nebula, no errors are given in \citet{esteban1998}
for the abundances deduced. For comparison, the solar values from AGS05 and
\citet{GS98} (hereafter GS98) are listed in the last two rows, respectively.

\section{oxygen and neon abundances of PNe and of H~{\sc ii} regions}

The data collected from the current survey corroborate the previous findings.
Firstly, a linear regression of the mean abundances of neon and oxygen
tabulated in Table~\ref{tab01} yields slopes of 1.21$\pm$0.03 and 1.14$\pm$0.01
with correlation coefficients of 0.98 and 0.89, for samples of PNe and \hiir\,
respectively. The slopes agree well with the previous results described above.
Secondly, it is clear that for both samples the Ne/O ratio increases with
increasing oxygen abundance, with a linear correlation coefficient larger than
0.75. Also in both types of objects, we see a discontinuity in the overall
positive correlation. Remarkably, this discontinuity occurs between Sextans~B
and NGC~3109 for both PNe and \hiir. In the above regression analysis, we have
excluded the PNe data point of Sextans~A, which contains only one object with
an abnormally low Ne/O ratio, probably caused by measurement errors.

We should emphasize that the positive correlation is possibly not introduced by
the hardening of spectra in low metallicity nebulae. We have constructed a set
of eight ideal photo-ionization models using the code CLOUDY (Ferland et al.
1998) with typical input parameters for PNe, varying only the oxygen abundance
from 8.7 to $7.0$ or the effective temperature of ionizing star from 75\,000 to
200\,000~K. The derived Ne/O ratios based on the output line intensities and
ICFs are nealy constant for all the models, with the largest variation only
about 10\%. This is reasonable given the similarities of ionization potentials
of oxygen and neon. To obtain an estimate of systematic errors involved in the
individual samples studied in this letter, we used our methods to recalculate
the Ne/O ratios and oxygen abundance for several samples based on the line
intensities published in the original papers. Our new O/H and Ne/O ratios are
consistent with the published values and no evident bias is found.

The positive correlation between Ne/O and O/H observed in \hiir\, indicates a
different enrichment history of neon and oxygen in the ISM. That is, the
enhancement of neon lags behind oxygen, coinciding with the current theory of
nucleosynthesis of MS. Kobayashi et al. (2006, K06) have calculated the
evolution of heavy-element abundances, adopting their own new nucleosynthesis
yields. Table~3 of K06 gives yields of Type~II supernovae and hypernovae
integrated over a Salpeter initial mass function (IMF) for different
metallicities.  The average oxygen abundance for Sextans A determined from
\hiir\, is 7.47, corresponds to a metallicity $Z\sim0.001$, at which K06
predict a Ne/O ratio of 0.17, consistent with the observed average value in
this galaxy. At solar metallicity, the K06 predict a Ne/O ratio of 0.28, again
close to what observed in the Orion Nebula and in M~51.  The agreement between
the observations and the theoretical predictions of K06 is remarkable. In this
comparison we have however neglected the contributions of SN\,Ia to the
evolution of O, Ne, and S, which are not all negligible.


The case of PNe is more complex. As described earlier, oxygen can be either
manufactured or destroyed during the late evolutionary stages of LIMS. From
Table~\ref{tab01} and more obviously from Fig.~\ref{fig02}, we find that for
four relatively metal-rich and massive galaxies, the Milky Way, LMC, SMC and
NGC 6822, oxygen and neon abundances measured in PNe are almost identical to
those found in \hiir. For the less massive and more metal-poor galaxies,
NGC~3109, Sextans~B and Sextans~A, the oxygen and neon abundances obtained for
PNe are significantly higher than those determined for \hiir, and the
differences increase as the metallicity decreases. The trend is similar for
both oxygen and neon, leading to small differences in the Ne/O abundance ratios
at all metallicities. The only exception is the neon abundance of the single PN
in Sextans~A, which is probably erroneous due to observational uncertainties. 

It is clear from the above result that the synthesis of oxygen and neon in LIMS
occurs only at low metallicities, and that the amount of oxygen and neon
synthesized seems to be comparable. When oxygen abundance reaches about 8.0 or
higher, the manufacture of oxygen and neon in LIMS diminishes. That
the yield of oxygen increases in low metallicity environments has also been
found previously by \citet{LD06} and is supported by theoretical calculations
of \citet{marigo2001}. The latter author shows that the net yield of $^{16}$O
is positive for stars of initial masses between $\sim0.8M_\odot$ and
$\sim3.5M_\odot$, thanks to the dredge-up events during the thermal-pulse AGB
(TP-AGB) phase, and that the yield increases with decreasing metallicities due
to higher efficiency and longer durations of the TP-AGB phase. 

We emphasize that although PNe in NGC~6822 have mean oxygen and neon abundances
comparable to PNe in Sextans~A, B and NGC~3109, PNe in NGC~6822 may have
experienced a different chemical enrichment history compared to those in the
latter three galaxies. It seems that oxygen and neon are not enhanced in PNe of
NGC~6822 but both elements are significantly enhanced in PNe of the latter
three galaxies. Therefore, if PNe of a galaxy are measured to have an oxygen
abundance close to 8.0, this abundance value can either result from a
progenitor abundance lower than 7.8 but then enhanced to the observed value
during the late evolutionary stages of the progenitor star, as in the case of
Sextans~B. The abundance can however reflect the true metallicity of the ISM
from which the PN was formed, as in the case of NGC~6822.

On the other hand, we find that oxygen destruction in the progenitor stars of
PNe is insignificant and does not affect the mean oxygen abundance of PNe,
though this phenomenon has been observed occasionally in some PNe evolved from
massive progenitors. This is consistent with the current theoretical
predictions. Model calculations by \citet{karakas2003} show that oxygen
destruction takes place only in stars more massive than 5\,$M_\odot$ of initial
metallicities from $Z = 0.004$ to 0.008 or in stars of initial masses between 6
and 6.5~$M_\odot$ at solar metallicity. What's more, in the former case, the
star will also exhibit high N/O ratios, inconsistent with the observations.
Clearly, oxygen destruction occurs only rarely and hardly affects the mean
oxygen abundance of PNe in nearby galaxies. For more distant galaxies, the
effect can be more significant as the observations may be biased towards PNe
descended from more massive stars in which oxygen destruction can be more
important. 

The consistency between Ne/O ratios observed in PNe and in \hiir\, of the same
host galaxies (excluding Sextans A), as shown in Fig.~\ref{fig02}, strongly
suggests that the Ne/O ratio is not altered by the late-stage evolution of
LIMS. That means, oxygen and neon are either hardly modified or altered by
comparable amounts in LIMS. The higher neon and oxygen abundances of PNe
compared to those of \hiir\ in NGC~3109 and the Sextans galaxies seem to
indicate the occurrence of the latter scenario in metal-poor environments. In
galaxies of higher metallicities, such as the LMC and the Galaxy, there is no
evidence of enhancement of oxygen and neon.

It is beyond the expectation of the current theory of nucleosynthesis of LIMS
that neon is enriched by amounts comparable to those of oxygen at low
metallicities. Detailed stellar evolutionary calculations for compositions
appropriate to the Galaxy and LMC by \cite{KL2003} have shown that neon
production becomes significant only in a narrow mass range, from about 2 to
4\,$M_\odot$ (lower for lower metallicities), where $^{22}$Ne is synthesized by
two $\alpha$-captures onto $^{14}$N in sufficient quantities to affect the
total neon abundance by more than 20\%. \cite{marigo2003} also suggest a
sizeable Ne production in intermediate-mass stars of the LMC composition.
Although some oxygen enrichment is expected to occur in a similar mass range,
the production of the two $\alpha$-elements proceeds via different channels.
$^{16}$O is synthesized mainly from $^{12}$C via $\alpha$-capture reactions and
possibly to a minor degree from $^{13}$C with neutrons as by-products, whereas
$^{22}$Ne is mainly synthesized from $^{14}$N. As such it is difficult to
understand why oxygen and neon are enhanced by comparable amounts. More
detailed theoretical calculations are highly desired. 

\begin{table}
\caption{Oxygen and neon abundances on a logarithmic scale where H = 12 and
Ne/O ratios for PNe and H~{\sc ii} regions. Numbers in parentheses are the
standard errors of the mean (In cases where the sample contains only one object,
the measurement errors are given instead).}
\label{tab01}
\begin{center}
\begin{tabular}{l c c  c c l}
\hline
         & N(obj) & O/H & Ne/H & Ne/O  & refs\\
\noalign{\smallskip}
\hline
PNe                \\
\noalign{\smallskip}
Leo A     &  1 & 7.30(0.05) & 6.38(0.11) & 0.12(0.03) & 1 \\
Sextans A &  1 & 8.00(0.10) & 6.70(0.20) & 0.05(0.03) & 2\\  
Sextans~B &  5 & 7.96(0.15) & 7.24(0.15) & 0.20(0.04) & 2 \\  
NGC~3109  &  6 & 8.16(0.19) & 7.24(0.42) & 0.14(0.03) & 3 \\
NGC~6822  &  6 & 8.01(0.13) & 7.32(0.35) & 0.20(0.03) & 4 \\
SMC       & 37 & 8.10(0.05) & 7.27(0.08) & 0.15(0.01) & 5 \\
LMC       &120 & 8.38(0.03) & 7.65(0.04) & 0.18(0.01) & 5 \\
M 31      & 12 & 8.40(0.09) & 7.65(0.11) & 0.19(0.01) & 6 \\
Gal. Halo &  9 & 7.99(0.07) & 7.10(0.16) & 0.20(0.08) & 7 \\
Gal. Disk & 58 & 8.60(0.02) & 7.99(0.04) & 0.24(0.01) & 8 \\
Gal. Bulge & 25 & 8.70(0.03) & 8.13(0.04) & 0.25(0.02) & 8 \\
\noalign{\smallskip}
\hline
H~{\sc ii} regions \\
\noalign{\smallskip}
Sextans~A &  4 & 7.47(0.12) & 6.68(0.12) & 0.16(0.02) & 2 \\  
Sextans~B &  3 & 7.63(0.03) & 6.83(0.17) & 0.18(0.05) & 2 \\  
NGC~3109  & 10 & 7.77(0.07) & 6.84(0.10) & 0.13(0.01) & 3 \\
dIrr      & 27 & 7.84(0.04) & 7.06(0.04) & 0.17(0.01) & 9 \\ 
BCDs      & 53 & 7.91(0.03) & 7.15(0.03) & 0.18(0.01) &10 \\
SMC       &  6 & 8.03(0.06) & 7.27(0.20) & 0.17(0.11) &11 \\  
NGC~6822  &  2 & 8.05(0.05) & 7.31(0.04) & 0.18(0.01) &12 \\
NGC~5253  &  4 & 8.24(0.04) & 7.53(0.04) & 0.19(0.01) &13 \\
M 33      &  6 & 8.27(0.02) & 7.57(0.02) & 0.20(0.01) &14 \\
LMC       &  4 & 8.35(0.06) & 7.61(0.05) & 0.18(0.03) &11 \\
M 51      &  7 & 8.58(0.03) & 7.91(0.09) & 0.24(0.04) &15 \\
Orion     &  1 & 8.60       & 8.00       & 0.25       &16 \\
\noalign{\smallskip}
\hline
\noalign{\smallskip}
Solar&     & 8.66       & 7.84       & 0.15       & 17\\
Solar&     & 8.83       & 8.08       & 0.18       & 18\\
\hline
\end{tabular}
\begin{list}{}{}
\item [References:]

[1] \citet{vanzee2006b}; [2] \citet{magrini2005}; [3] \citet{pena2007}; [4]
\citet{richer2007}; [5] \citet{LD06}; [6] Jacoby \& Ciardullo (1999, excluding
PN F57, 455 and 470); [7] \cite{howard1997}; [8] Wang \& Liu (2007, from CELs);
[9] \citet{vanzee2006a}; [10] \citet{guseva2007}; [11] \citet{RD1990}; [12]
\citet{peimbert2005}; [13] \citet{lopez-sanchez2007}; [14]
\citet{crockett2006}; [15] \citet{bresolin2004}; [16] \cite{simpson2004}; [17]
\cite{asplund2005}; [18] \cite{GS98}.

\end{list}
\end{center}
\end{table}

\section{The solar neon abundance}

The uncertainty of solar neon abundance is intrinsically large. It is generally
determined by observing the outer regions (such as corona) of the Sun, and
relies on the relative abundance of neon with respect to oxygen or magnesium
assuming that relative abundance ratios in the corona are equal to the
photospheric values. From the ratio of Ne/O (Ne/Mg) and the photospheric
abundance of oxygen (magnesium), one can deduce the photospheric absolute
abundances of neon. However, this method suffers from large potential
uncertainties when applying corrections to account for the first ionization
potential (FIP) effects, especially when employing the ratio of Ne/Mg, as
magnesium is a low FIP element and oxygen and neon are not. The current best
estimates of the solar Ne/O ratio and neon abundance obtained from this method
are 0.15 and 7.84, respectively (cf. AGS05). However, this very
low Ne/O ratio, when combined with the much reduced photospheric oxygen
abundance obtained from the application of a time-dependent 3D hydrodynamical
model of the solar atmosphere (AGS05), runs into serious conflict with the
current best solar model. For example, \citet{bahcall2005} showed that
helioseismological measurements require a Ne/O ratio as high as $\sim0.4$.
Interestingly, recent observations and analyses of a sample of nearby
solar-like stars by \citet{drake05} also support a high Ne/O ratio of about
0.4.

Recently, \citet{basu2008} have examined possibilities which could reconcile
the large discrepancies between the Sun and models constructed with the AGS05
abundances. They show that the discrepancies are too large to be accounted for
by potential uncertainties in the opacity calculations. Increasing the
diffusion does not help much either. Raising the neon abundance is probably the
best choice to increase the opacity in order to compensate for the reduction in
oxygen abundance yet not introduce other inconsistencies. According to their
calculations, raising the neon abundance by a factor of 4 or 2.5 (the latter
value is obtained by increasing the abundances of C, N and O by 1$\sigma$
uncertainties), which corresponds to a Ne/O ratio of 0.6 and 0.4, respectively,
could compensate for the reduction in the oxygen abundance. Although, these
values appear too high to be supported by the available measurements, even
though it is quite possible that there are other factors affecting the
analysis, their calculations do suggest raising the neon abundance, if the
solar photospheric abundances of other abundant second-row elements recommended
by AGS05 are indeed correct. 

\begin{figure}
\centering
\epsfig{file=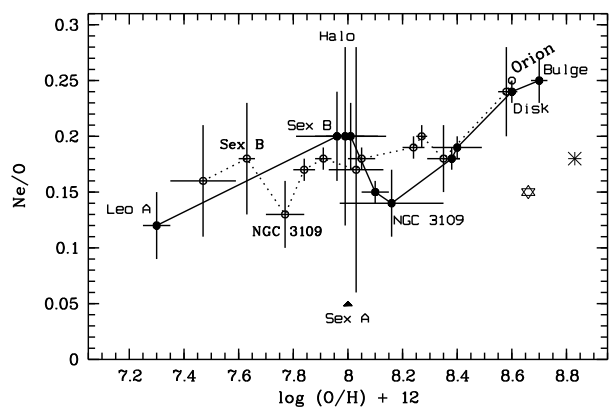, width=8.0cm, bbllx=48,
bblly=348,bburx=442, bbury=616,clip=,angle=0}
\caption{The mean Ne/O abundance ratios plotted against the average oxygen
abundances derived from \hiir\,(open symbols) and from PNe (filled symbols) in
nearby galaxies. The star and asterisk stand for the solar values given by
AGS05 and by GS98, respectively. Names of several sources are marked. Different
types of sources are linked by lines to aid visualization.} 
\label{fig01}
\end{figure}

\begin{figure}
\centering
\epsfig{file=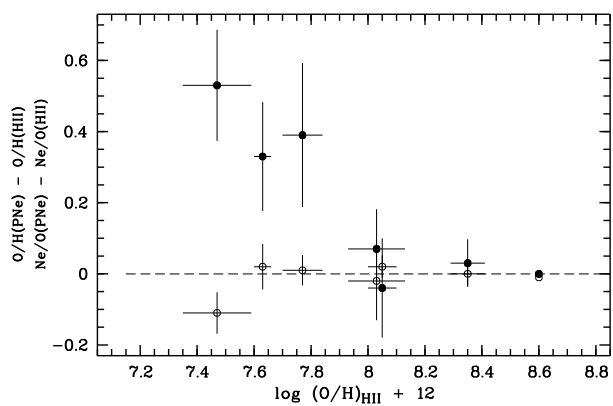, width=8.0cm, bbllx=98,
bblly=348,bburx=497, bbury=616,clip=,angle=0}
\caption{The differences between oxygen abundances (filled circles) and the
Ne/O ratios (open circles) of PNe and of \hiir\, are plotted against the mean
oxygen abundances of \hiir\, in 7 galaxies. From left to right, they are
Sextans~A, Sextans~B, NGC~3109, SMC, NGC~6822, LMC and the Galaxy, respectively.}
\label{fig02}
\end{figure}

The Ne/O ratios observed in PNe and \hiir\, provide important clues to the
solar Ne/O ratio. Fig.~\ref{fig01} shows that the solar Ne/O abundance ratios
(recommended by GS98 and by AGS05) are definitely too low compared to the mean
ratio of $\sim0.25$ measured in Galactic disk PNe and \hiir, both having an
average oxygen abundance comparable to the Sun. As concluded in the last
section, there is no clear evidence of modification of Ne/O ratios during the
late evolutionary stages of LIMS. Therefore they should yield a Ne/O ratio
comparable to the solar value. Hence, we strongly suggest that the current
solar Ne/O value, and consequently also the absolute neon abundance, should be
raised by a factor of 1.7, i.e. 0.22~dex, from the AGS05 values, or by a factor
of 1.4, i.e. 0.14~dex, from the GS98 values.

We note that the depletion of oxygen onto dust grain is insignificant at solar
metallicity and will not change the above conclusion. \citet{simondiaz2006}
have compared the oxygen abundance obtained from a detailed and fully
consistent spectroscopic analysis of the group of B stars associated with the
Orion Nebula with recent nebular gas-phase results. They found that they are in
good agreement with each other and therefore the dust depletion is quite small
($\la0.02$~dex) at the metallicity of Orion. 

Rubin et al. (2008) have determined the Ne/S ratios for 25 low-metallicity,
high-ionization \hiir\, in the local group spiral galaxy M33, based on {\it
Spitzer} data by sampling the dominant ionization states of Ne (Ne$^{+}$,
Ne$^{++}$) and S (S$^{++}$, S$^{3+}$). Combined with other results (cf. their
Figs. 11 and 12), the authors regarded their estimated total Ne/S ratio to be
reliable. Their derived mean Ne/S ratios range from 10.1 to 16.3, much higher
then the solar value of $\sim5$ (AGS05), and consistent with the solar Ne/S
ratio if Ne abundance alone is raised to 8.29, as suggested by
\citet{bahcall2005}. The sharp contrast between their mean Ne/S ratio and the
solar ratio strongly supports our conclusion here that the solar neon abundance
is currently largely underestimated. Rubin et al. (2008) also noted the
possibility that the true Ne/S ratio may be less for lower metallicity
galaxies, in accordance with our other conclusion that the Ne/O ratio is less
in lower metallicity galaxies, due to a delay of neon production at low
metallicity.

\section{Conclusion}

In this letter, we present a critical literature survey of recent measurements
of oxygen and neon abundances of PNe and of \hiir\, in the Milky Way and other
nearby galaxies. We find that there is significant oxygen and neon production
in LIMS at metallicities lower than the SMC, but not at higher metallicities.
We find that oxygen destruction is probably insignificant. We show that neon
and oxygen are probably enhanced by the same amounts in PNe at low
metallicities, a result not predicted by the current theory of nucleosynthesis
for LIMS. 

We find that the Ne/O ratio increases with increasing oxygen abundance in PNe
and in \hiir, suggesting a different enrichment history of neon and oxygen in
the ISM and thus probably different production mechanisms of these two
$\alpha$-elements in massive stars, as predicted by current theoretical
calculations. 

Both PNe and \hiir\, in the Galactic disk yield a consistent Ne/O ratio of
0.25, higher than the solar value of 0.18 (GS98) or 0.15 (AGS05). We suggest
that the solar Ne/O ratio and the absolute neon abundance need to be revised
upwards by about 0.22~dex from the values of AGS05 or by 0.14~dex from those of
GS98. A recent discussion based on the Ne/S ratios measured in \hiir\, in
nearby galaxies by \citet{rubin2008} also points to an upwards revision to the
solar neon abundance. 

$$\\$$
The work is partially supported by Grant \#10325312 of the National Natural
Science Foundation of China. We would also like to thank the referee for 
useful comments.

\end{document}